# Multiple flat bands and localized states in photonic super-Kagome lattices

LIMIN SONG,[1] SHENYI GAO,[1] JINA MA,[2] LIQIN TANG,[1,3,*] DAOHONG SONG,[1,3,*] YIGANG LI,[1] AND ZHIGANG CHEN[1,3,*]

[1] *The MOE Key Laboratory of Weak-Light Nonlinear Photonics, TEDA Applied Physics Institute and School of Physics, Nankai University, Tianjin 300457, China*
[2] *School of Physics and New Energy, Xuzhou University of Technology, Xuzhou 221018, China*
[3] *Collaborative Innovation Center of Extreme Optics, Shanxi University, Taiyuan, Shanxi 030006, China*
*\*Corresponding author: tanya@nankai.edu.cn; songdaohong@nankai.edu.cn; zgchen@nankai.edu.cn*

**We demonstrate multiple flat bands and compact localized states (CLSs) in a photonic super-Kagome lattice (SKL) that exhibits coexistence of singular and nonsingular flat bands within its unique band structure. Specifically, we find that the upper two flat bands of an SKL are singular—characterized by singularities due to band touching with their neighboring dispersive bands at the Brillouin zone center. Conversely, the lower three degenerate flat bands are nonsingular, and remain spectrally isolated from other dispersive bands. The existence of such two distinct types of flat bands is experimentally demonstrated by observing stable evolution of the CLSs with various geometrical shapes in a laser-written SKL. We also discuss the classification of the flat bands in momentum space, using band-touching singularities of the Bloch wave functions. Furthermore, we validate this classification in real space based on unit cell occupancy of the CLSs in a single SKL plaquette. These results may provide insights for the study of flatband transport, dynamics, and nontrivial topological phenomena in other relevant systems.**

Flatband lattices are characterized by an energy band structure in which at least one band is completely dispersionless in the whole Brillouin zone [1, 2]. Initially proposed in condensed matter physics [3, 4], flatband lattices have gained considerable attention in the realm of electronics, acoustics, and photonics (see recent reviews [5-8] and references therein). Since the early pioneering experimental works [9-12] in photonics, recent flatband research interests have extended to creating light localization in non-Hermitian systems [13] and moiré lattices [14, 15], promoting quantum network transfer and storage [16], and realizing flat bands in synthetic dimensions [17-19]. Meanwhile, flat bands also play important roles in for example Abelian and non-Abelian Thouless pumping [20, 21], quantum error filtering [22], bimorphic Floquet topological insulators [23], and the interaction between localized electrons and extended photons [24].

Flatband lattices can support a large number of entirely degenerate eigenstates, namely, the compact localized states (CLSs) [1, 2, 25], which only occupy some finite lattice sites with completely vanishing amplitude elsewhere. Conventionally, flat bands were grouped into three distinct types: "symmetry-protected", "accidental", and "topologically protected" flat bands [2]. However, these classifications may not effectively characterize features of all flat bands, especially those involve flatband touching/crossing with other energy bands. From the viewpoint of real-space state counting, the band touching gives rise to an additional number of linearly independent localized states with flatband energy and is usually mediated by nontrivial real-space topology [1, 26, 27]. Meanwhile, such band touching also gives rise to the singularity/discontinuity of the flatband Bloch wave function (BWF), which has been proposed as an effective way to classify singular and nonsingular flat bands [28]. Moreover, flatband-touching singularity mediates many intriguing physical effects, including for example nontrivial flatband line states and noncontractible loop states [27-32], nonzero quantum distance and anomalous Landau levels [33], as well as novel bulk-interface correspondence [34].

In this Letter, we present theoretical and experimental studies on multiple flat bands and relevant CLSs in super-Kagome lattices (SKLs), unveiling their flatband features in both real and momentum spaces. Specifically, we classify flat bands based on their supporting CLSs in real space and corresponding momentum-space BWFs. In momentum space, a flat band can either be singular or nonsingular, depending on the presence or absence of a BWF singularity [28]. In real space, we define the counterparts of CLSs (and hence flat bands) in two-dimensional settings based on an integer number U, which characterizes the occupied unit cells in one-dimensional settings [25]. We find these classifications can be simultaneously manifested in the analytical expression of the BWFs: each component of a BWF is a finite sum of the Bloch phases, and its distribution in real space can be obtained through an inverse Fourier transformation [28, 35]. Furthermore, we experimentally excite the flatband CLSs with different propagation constants at the input of a photonic SKL, and demonstrate their diffractionless propagation, confirming the existence of multiple flat bands.

Inspired by the general method to construct superlattices [36-38], we insert another waveguide (green sites) halfway into each bond of the typical Kagome model (orange sites) to construct an SKL, as sketched in Fig. 1(a). The SKL, therefore, consists of nine sublattices $(S_1, S_2, \cdots, S_9)$ in each unit cell, highlighted by the dashed rhombus [Fig. 1(a)]. An experimentally realized finite SKL of evanescently coupled waveguides (with 34 μm spacing) is shown in Fig. 1(b), established by the continuous-wave laser-writing technique in a nonlinear crystal [29, 39, 40]. Under the tight-binding approximation with only the nearest-neighbor coupling considered, the Bloch Hamiltonian of an infinite SKL takes the form as

$$H = t \begin{pmatrix} 0 & 1 & 1 & 0 & 0 & 0 & e^{ik_2} & 0 & e^{-ik_3} \\ 1 & 0 & 1 & 1 & 0 & 0 & 0 & 0 & e^{-ik_3} \\ 1 & 1 & 0 & 1 & 0 & 0 & e^{ik_2} & 0 & 0 \\ 0 & 1 & 1 & 0 & 1 & 1 & 0 & 0 & 0 \\ 0 & 0 & 0 & 1 & 0 & 1 & 1 & 1 & 0 \\ 0 & 0 & 0 & 1 & 1 & 0 & 0 & 1 & 1 \\ e^{-ik_2} & 0 & e^{-ik_2} & 0 & 1 & 0 & 0 & 1 & 0 \\ 0 & 0 & 0 & 0 & 1 & 1 & 1 & 0 & 1 \\ e^{ik_3} & e^{ik_3} & 0 & 0 & 0 & 1 & 0 & 1 & 0 \end{pmatrix}, \quad (1)$$

where $t$ is the uniform nearest-neighbor coupling strength, $k_i = \mathbf{k} \cdot \mathbf{a}_i$ $(i = 1,2,3)$, $\mathbf{k} = (k_x, k_y)$ is the Bloch momentum, $\mathbf{a}_1 = a(1,0)$, $\mathbf{a}_2 = a(1/2, \sqrt{3}/2)$, and $\mathbf{a}_3 = a(1/2, -\sqrt{3}/2)$ are the lattice vectors shown as the black arrows in Fig. 1(a). We assume $a = 1$ and $t = 1$ for simplicity.

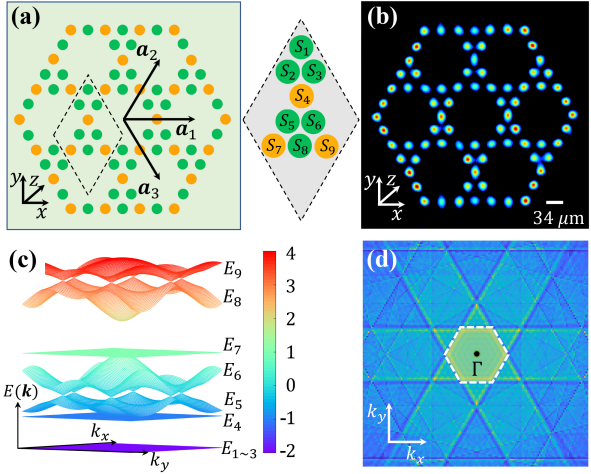

**Fig. 1.** (a) Schematic diagram of an SKL, with nine basis sites per unit cell (marked as $S_1, S_2, \cdots, S_9$ in the right inset). $\mathbf{a}_{1,2,3}$ denote the three lattice vectors. (b) An experimentally established photonic SKL corresponding to (a). (c) Calculated band structure of an infinite SKL under the tight-binding approximation. There are four dispersive bands and five flat bands, where the bottom flat bands at $E_{1,2,3} = -2$ are nonsingular and triply degenerate, and other two flat bands are singular, located at $E_4 = -1$ and $E_7 = 1$. (d) Calculated Brillouin zone (BZ) spectrum of the SKL. The dashed hexagon marks the first BZ and Γ is its center.

By solving the Bloch eigenvalue problem $H|v_n(\mathbf{k})\rangle = E_n|v_n(\mathbf{k})\rangle$, where $n = 1,2,\cdots,9$ counts the energy bands, one can obtain the band structure $E_n$ (or the propagation constant) of the lattice, as shown in Fig. 1(c). The corresponding calculated BZ spectrum is displayed in Fig. 1(d). One can find that there are seven energy bands with four dispersive bands and three flat ones. It is worth noting that the bottom isolated flat band is triply degenerate at $E = -2$. Therefore, we have five flat bands in total. From bottom to top, these energy bands are labeled from $E_1$ to $E_9$. The flat band $E_4$ ($E_7$) touches its neighbor upper (lower) dispersive bands $E_5$ ($E_6$) at the Γ point in the first BZ. Besides, there are also 12 pairs of pseudospin-1/2 Dirac cones intersected by the two groups of paired dispersive bands. In the following, we focus on these flat bands and the associated CLSs.

In order to reveal whether these flat bands have singular properties, let us first consider the triply degenerate flat bands at $E_{1,2,3} = -2$. The normalized flatband BWFs can be obtained as
$$|v_1\rangle = \alpha_1^{-1}(1, -1, -1, 2, -1, -1, 0, 1, 0)^{\mathrm{T}}, \quad (2)$$
$$|v_2\rangle = \alpha_2^{-1}(1 - e^{ik_2}, e^{ik_2} - 1, -1 - e^{ik_2}, 2, -2, 0, 2, 0, 0)^{\mathrm{T}}, \quad (3)$$
$$|v_3\rangle = \alpha_3^{-1}(1 - e^{-ik_3}, -1 - e^{-ik_3}, e^{-ik_3} - 1, 2, 0, -2, 0, 0, 2)^{\mathrm{T}}, \quad (4)$$
where $\alpha_1 = \sqrt{10}$, $\alpha_2 = (18 - 2\cos k_2)^{1/2}$, and $\alpha_3 = (18 - 2\cos k_3)^{1/2}$ are the normalization factors. For the BWF $|v_1\rangle$, the zero components indicate the flatband states never populate $S_7$ and $S_9$ sublattices, and the amplitudes at $S_1$ and $S_8$ sublattices are 1, and those at $S_{2,3,5,6}$ are $-1$, and at $S_4$ is 2. The real-space profile of one localized state (CLS-1), corresponding to $|v_1\rangle$, is shown in Fig. 2(a). Such a state is only within one unit cell $\mathbf{R} = \mathbf{0}$ ($\mathbf{R}$ is a selected lattice vector indicating the relative locations of the unit cells), and therefore the corresponding flat band can be classified as $U = 1$ class in real space. A recent study has shown that the completeness of the CLS is encapsulated in the normalization factor $\alpha$ [28]. Because $\alpha_1$ is a constant and not zero, its BWF has no singularities and corresponding flat band is nonsingular in momentum space. Next, we obtain the real-space profiles of the CLS-2 and CLS-3 from the BWFs $|v_2\rangle$ and $|v_3\rangle$, shown in Figs. 2(b) and 2(c), respectively. These two CLSs individually populate two unit cells ($\mathbf{R} = \mathbf{0}$ and $\mathbf{R} = -\mathbf{a}_2$ for CLS-2, $\mathbf{R} = \mathbf{0}$ and $\mathbf{R} = \mathbf{a}_3$ for CLS-3) and the corresponding flat bands can be classified as $U = 2$. By checking $\alpha_{2,3}$, we find also that there are no singularities and therefore the isolated triply degenerate flat bands are nonsingular. Such non-singularity also manifests in their real-space classifications. For example, the flat band hosting basic CLSs in $U = 1$ class is obviously nonsingular because one cannot find singularity in the relevant BWF. Such a kind of CLSs can naturally form an orthogonal set, promising for constructing the flatband quantum scar [41]. As for the CLSs in $U = 2$ class, despite that BWFs $\alpha_2|v_2\rangle$ and $\alpha_3|v_3\rangle$ have $\mathbf{k}$-dependent components, the $\mathbf{k}$-independent components contribute an inescapable effort to the non-singularity of the flat bands. In fact, the CLS-2 and CLS-3 can be decomposed into two smaller CLSs: the CLS-1 and its neighbor rotation. Reversely, more complicated CLSs can be constructed by linear combinations of the simple CLSs. Here we show some typical examples for $E = -2$, the rhombic CLS-6 [Fig. 2(f)] in $U = 4$, the shield-like CLS-7 [Fig. 2(g)] in $U = 4$, and the armchair-like CLS-8 [Fig. 2(h)] in $U = 3$. For the isolated degenerate flat bands at $E = -2$, the smallest possible CLS having a lower $U$ class dominates whether it is singular or not, although there are many choices of the CLS sets to span the flat bands.

Next, we discuss the other two flat bands $E_4$ and $E_7$ touching the dispersive bands. The normalized BWF for $E_4 = -1$ reads
$$|v_4\rangle = \alpha_4^{-1}(1 - e^{ik_1}, e^{ik_1} - e^{ik_2}, e^{ik_2} - 1, 0, e^{ik_3} - e^{ik_1},$$
$$1 - e^{ik_3}, 0, e^{ik_1} - 1, 0)^{\mathrm{T}}, \quad (5)$$
and that for $E_7 = 1$ reads
$$|v_5\rangle = \alpha_5^{-1}(e^{ik_2} - e^{-ik_3}, e^{ik_2 - ik_3} - e^{ik_2}, e^{-ik_3} - e^{ik_2 - ik_3},$$
$$2e^{-ik_3} - 2e^{ik_2}, 1 - e^{ik_2}, e^{-ik_3} - 1, 2 - 2e^{-ik_3}, e^{ik_2} - e^{-ik_3},$$
$$2e^{ik_2} - 2)^{\mathrm{T}}, \quad (6)$$
The normalization factors are $\alpha_4 = [2(6 - 3\cos k_1 - \cos k_2 - 2\cos k_3)]^{1/2}$ and $\alpha_5 = [12(3 - \cos k_1 - \cos k_2 - \cos k_3)]^{1/2}$. The immovable discontinuity at the Γ point, i.e., $\mathbf{k} = (0,0)$, of the BZ reveals that they belong to the classification of singular flat bands [8, 28]. The analytical BWF Eq. (5) and the corresponding mode distribution Fig. 2(d) clearly illustrate that the optical field of the CLS-4 is absent from the $S_{4,7,9}$ sites and strongly localized at the other sites; while the CLS-5 described by Eq. (6) occupy all kinds of sublattices, as shown in Fig. 2(e). Besides, we can see that both of the two CLSs of the singular flat bands can be classified as $U = 4$, the maximum unit cell occupancy around a single (hexagonal) plaquette. Examples like such structures range from Kagome [11, 26, 28] and Lieb [9, 10, 29] to decorated honeycomb [30, 31] and fractal-like [32, 42, 43] lattices, indicating a remarkable feature of singular flatband touching. For particular lattice geometries, the maximum unit cell occupancy may relate to spatial dimensionality, structural details, and certain symmetries [35]. Within the scope of this work, we only make the above flatband classifications concerning $U$ for practical reasons.

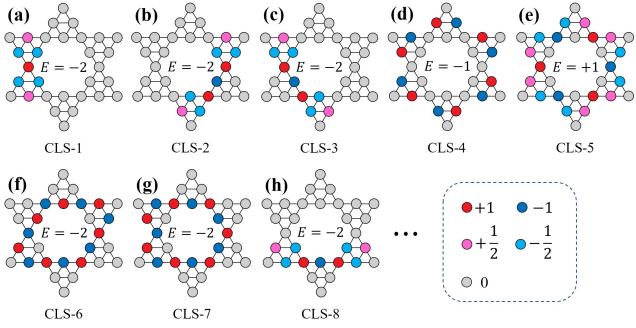

**Fig. 2.** Real-space amplitude and phase distributions of the CLSs of singular and nonsingular flat bands based on a discrete model. These CLSs correspond to flatband energies $E = -2$ in (a-c) and (f-h), $E = -1$ in (d), and $E = 1$ in (e). The values for different colored lattice sites are $+1$ (red), $-1$ (dark blue), $+1/2$ (magenta), $-1/2$ (peacock blue), and $0$ (gray). $+/-$ represent phases $0/\pi$.

We discuss the formation mechanism of the two distinct types of flat bands: two singular flat bands touching with dispersive bands and three degenerate nonsingular flat bands. The SKL is constructed by symmetrically decorating the upward and downward triangles of the underlying Kagome lattice—a commonly known model exhibiting strong geometrical frustration and with a singular flatband touching [26, 28]. In addition, each individual decoration with respect to the upward or downward triangle is analogous to a fractal self-similar process, and the resulting lattice is the emerging fractal-like lattice with the unit cell being the first generation of Sierpinski gasket [42-44]. As we have seen, the complex construction of the symmetric lattice facilitates the existence of the CLSs caused by local destructive interference, which allows for a mixture of various localized states even with unequal amplitudes. Thus, the SKL somehow directly inherits the dual properties of geometrical frustration and fractal self-similarity, which are the key ingredients accounting for the flatband singularity and flatband degeneracy.

To demonstrate the flatband CLSs experimentally, we employ a laser-written photonic SKL (Fig. 1b) as the platform for their implementation. The experimental setup is similar to that in our previous work [29]. We focus on the diversiform profiles of the CLSs, characteristic feature of multi-flatband systems. We first consider the CLSs with the same energy, such as the CLS-6, 7, 8 belonging to the bottom degenerate flat bands. Then, the CLSs for the middle two singular flat bands (the CLS-4 and CLS-5) are also demonstrated. To excite the CLSs, a probe beam with specific amplitude and phase distribution matching that of the calculated flatband states is launched into the SKL (see Fig. 2). A spatial light modulator (SLM) is used to precisely control the excitation position of the probe beam, as well as the amplitude and phase. For the initial alternating-phase excitations, the output fields of the corresponding CLSs at a propagation distance of 10 mm through the lattice are shown in Figs. 3(a1-a5). One can see that the overall shape of these CLSs remains intact after propagation, which demonstrates the existence of CLSs in SKL. By the same token, these observations clearly indicate the existence of the two distinct types of flat bands in the photonic SKL. In addition, stable transport of these complex CLSs under linear propagation may be useful for various applications, including but not limited to transporting images and codified information [7, 16, 29].

For the initial equal-phase excitations (no phase difference between sites), the relevant results are shown in Figs. 3(b1-b5). As expected, the input beam cannot be localized and the output patterns exhibit discrete diffraction with light energies escaping to other initially un-excited sites (not shown here, but the first row, Figs. 3(a1)-3(a5), can be used as a reference because the alternating-phase cases remain unchanged during propagation). Although these output patterns display asymmetry to some extent due to the anisotropic nonlinear crystal, the clear contrast between the alternating- and equal-phase cases unambiguously demonstrates the existence of the different types CLSs in SKLs. We note that the excitation of CLSs requires careful engineering in the phase distribution between lattice sites.

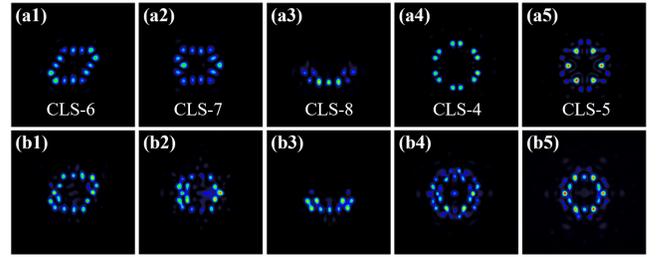

**Fig. 3.** Experimental demonstration of distinct CLSs with various geometrical shapes in a photonic SKL. (a1)-(a5) The output intensity patterns of the CLSs under alternating-phase excitations, corresponding to the initial amplitude-and-phase distributions and the exciting positions shown in Fig. 2. (a1)-(a3) belong to the triply-degenerate nonsingular flat bands at $E = -2$, and (a4) and (a5) belong to the singular flat bands at $E_4 = -1$ and $E_7 = 1$, respectively. From left to right, shown are the rhombic CLS-6, shield-like CLS-7, armchair-like CLS-8, necklace-like CLS-4, and crown-like CLS-5. The second row (b1)-(b5) has the same layout as the first row but under equal-phase excitations. Compact localization and noticeable discrete diffraction to other sites capture the features of the first and the second rows, respectively.

To corroborate the above experimental observations, we perform numerical simulations based on a continuous model by using the beam propagation method for even longer-distance propagation. In Figs. 4(a1)-4(a5), we show the photonic SKLs overlying with input beams to clearly show the corresponding excitation positions. These simulations are carried out under the same settings as the experimental parameters. One can see, the CLSs are still localized perfectly well during evolution under the alternating-phase excitations [Figs. 4(b1)-4(b5)], for a propagation distance three times than the experimental one. On the contrary, the equal-phase excitations diffract to more bulk sites [Figs. 4(c1)-4(c5)]. These CLSs are supported by the local symmetry of the lattice and stabilized by perfectly destructive interference. The key message through these simulations is highly consistent with that of the experimental observations in Fig. 3. These results illustrate that the two types of flat bands and the interesting CLSs with various geometries indeed reside in the SKL.

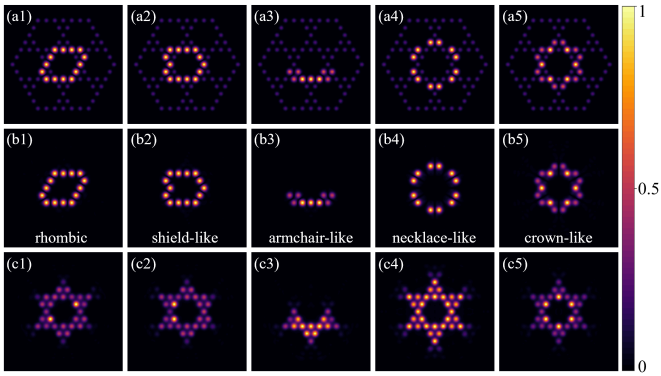

**Fig. 4.** Simulation results of the CLSs in a photonic SKL based on the continuous model. (a1-a5) Different input beams superimposed on the SKL, where the bright sites indicate the excited position and shape of the input beam, the background with relatively dark-sites show the SKL. (b1-b5) Numerical simulations of the distinct CLSs under alternating-phase excitations. The layout is the same as that for Figs. 3 (a1-a5), except that they are obtained at a longer propagation distance of 30 mm. (c1-c5) Same as (b1-b5) but under equal-phase excitation condition for a direct comparison—in the latter case, the patterns are not stationary as they deform during long-distance propagation.

In conclusion, we have reported the observation of CLSs in a photonic SKL. The SKL consists of multiple singular and nonsingular degenerated flat bands. We have classified these flat bands according to their singularities of the momentum-space BWFs and the occupancy of the real-space unit cells, and provided direct observation of the relevant CLSs. Furthermore, we have compared the CLSs supported by singular nondegenerate and nonsingular degenerate flat bands. We have highlighted the advantage of using higher-order-degenerate flat bands for achieve large-capacity image transmission. This work represents an ongoing effort towards creating artificial photonic flat bands with localized states and novel flatband features. The photonic SKL platform proposed here holds promise for exploring both fundamental principles and practical applications. For example, it can be applied to investigate momentum-space and real-space topology, as well as the interplay between these two distinct properties.

**Funding.** National Key R&D Program of China (2022YFA1404800); National Natural Science Foundation of China (12134006, 12274242 and 12374309); Natural Science Foundation of Tianjin (21JCYBJC00060, 21JCJQJC00050); 111 Project (B23045) in China.

**Disclosures**. The authors declare no conflicts of interest.

**Data availability.** Data underlying the results presented in this paper are not publicly available at this time but may be obtained from the authors upon reasonable request.